\documentclass[conference]{IEEEtran}
\usepackage{amsmath}
\usepackage{amssymb}
\usepackage{amsfonts}
\usepackage{graphicx}
\usepackage{epsfig}
\usepackage{subfigure}
\usepackage{psfrag}
\usepackage{cite}
\usepackage{latexsym}
\usepackage{url}
\usepackage{color}
\usepackage{hyperref}
\usepackage[hyphenbreaks]{breakurl}
\PassOptionsToPackage{bookmarks={false}}{hyperref}

\begin{document}
\title{Harnessing Self-Interference in Full-Duplex Relaying: An Analog Filter-and-Forward Approach
\author{\IEEEauthorblockN{Jie Xu\IEEEauthorrefmark{1}\IEEEauthorrefmark{2},
Lingjie Duan\IEEEauthorrefmark{2}, and Rui Zhang\IEEEauthorrefmark{3}
}
\IEEEauthorblockA{\IEEEauthorrefmark{1}School of Information Engineering, Guangdong University of Technology}
\IEEEauthorblockA{\IEEEauthorrefmark{2}Engineering Systems and Design Pillar, Singapore University of Technology and Design}
\IEEEauthorblockA{\IEEEauthorrefmark{3}Department of Electrical and Computer Engineering, National University of Singapore\\Email: jiexu.ustc@gmail.com, lingjie\_duan@sutd.edu.sg, elezhang@nus.edu.sg}
}}



\maketitle

\begin{abstract}
This paper studies a full-duplex filter-and-forward (FD-FF) relay system in frequency-selective channels. Conventionally, the loop-back signal at the FD relay is treated as harmful self-interference and needs to be significantly suppressed via both analog- and digital-domain cancellation. However, the performance of the conventional self-interference cancellation approach is fundamentally limited due to the quantization error induced by the analog-to-digital converter (ADC) with limited dynamic range. In this paper, we consider an analog filter-and-forward design to help avoid the quantization error, and surprisingly show that the maximum achievable rate of such an FD-FF relay system is in fact regardless of the loop-back channel at the FD relay. We characterize the maximum achievable rate of this channel by jointly optimizing the transmit power allocation over frequency at the source and the frequency response of the filter at the relay, subject to their individual power constraints. Although this problem is non-convex, we obtain its optimal solution by applying the Lagrange duality method. By simulations it is shown that the proposed joint source and relay optimization achieves rate gains over other heuristic designs, and is also advantageous over the conventional approach by cancelling the relay loop-back signal as self-interference, especially when the residual self-interference after cancellation is still significant.
\end{abstract}

\newtheorem{definition}{\underline{Definition}}[section]
\newtheorem{fact}{Fact}
\newtheorem{assumption}{Assumption}
\newtheorem{theorem}{\underline{Theorem}}[section]
\newtheorem{lemma}{\underline{Lemma}}[section]
\newtheorem{corollary}{\underline{Corollary}}[section]
\newtheorem{proposition}{\underline{Proposition}}[section]
\newtheorem{example}{\underline{Example}}[section]
\newtheorem{remark}{\underline{Remark}}[section]
\newtheorem{algorithm}{\underline{Algorithm}}[section]
\newcommand{\mv}[1]{\mbox{\boldmath{$ #1 $}}}
\setlength\abovedisplayskip{0.3pt}
\setlength\belowdisplayskip{0.3pt}

\section{Introduction}

Wireless relaying has been recognized as a promising solution to increase spectral efficiency and extend communication range in wireless networks, where relay nodes are deployed to help the information transmission from source nodes to destination nodes. Conventional relays often operate in half-duplex to transmit and receive at different time slot and/or over different frequency bands. With the recent advancements of full-duplex (FD) radios \cite{Sabharwal2014}, FD relays, which can simultaneously receive and forward information over the same frequency band, are attracting increasing attention to further improve the spectral efficiency in emerging wireless networks (see, e.g., \cite{Liu2015} and the references therein).

Due to the simultaneous transmission and reception, how to deal with the strong loop-back signal from the transmit to the receive antennas at FD relays becomes one of the key challenges to be tackled. In the existing literature, the loop-back signal is often treated as harmful self-interference and needs to be significantly suppressed beforehand (see, e.g., \cite{Ju2009,Riihonen2011,Krikidis2012,Day2012,Bharadia2014}). Towards this end, various self-interference mitigation techniques have been developed, which can be implemented in propagation-domain, analog-circuit-domain, digital-domain, etc. \cite{Sabharwal2014}. However, these techniques are difficult to perfectly cancel the strong self-interference. Specifically, for the most commonly employed self-interference cancellation in the digital domain, i.e., after the analog-to-digital converter (ADC), the limited dynamic range of the ADC induces significant quantization error that is not cancellable \cite{Sabharwal2014}.{\footnote{Under a given dynamic range of the ADC, the strong self-interference reduces the desirable signal power and relatively increases the quantization error.}} Such  residual self-interference will significantly degrade the system performance. Under the conventional approach and in the presence of residual self-interference, the authors in \cite{Ju2009,Riihonen2011,Krikidis2012,Day2012} have analyzed the information-theoretic achievable rate of FD relay channels, where the direct link from the source to the destination has been ignored for simplicity. In \cite{Bharadia2014}, the authors implemented an FD relay testbed in orthogonal frequency-division multiplexing (OFDM) systems, where the signal forwarded by the FD relay can be constructively combined at the destination with the signal transmitted from the source in the direct link; whereas self-interference is still required to be effectively cancelled at the FD relay.

\begin{figure}
\centering
 \epsfxsize=1\linewidth
    \includegraphics[width=8.2cm]{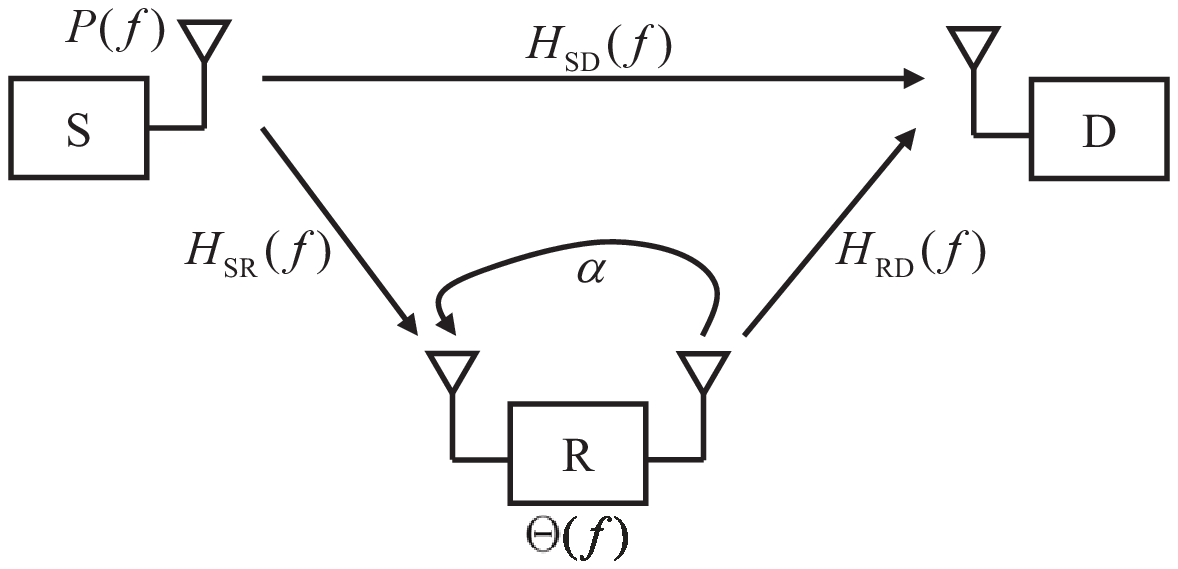}\vspace{-0em}
\caption{The system model of an FD relay R helping the information transmission from a source S to a destination D in frequency-selective channels.} \label{fig:1}\vspace{-1.5em}
\end{figure}

For FD relays, however, we observe that the loop-back signal is indeed a {\it delayed} version of the output signal of the relay processing which contains useful information signal of the source, and thus it should be utilized instead of being cancelled as self-interference as in the conventional approach. Under such a new approach, the capacity of the FD relay system under various channel setups and relay processing techniques still remains largely unknown, which motivates this paper.

We consider an FD relay channel as shown in Fig. \ref{fig:1}, where an FD relay node R operates in a filter-and-forward (FF) manner for helping the information transmission from a source node S to a corresponding destination node D. We consider that the FF processing at the FD relay is implemented in the analog domain (and yet digitally controlled), thus avoiding the quantization error induced by the ADC in the conventional approach of canceling the self-interference in digital domain. In our model, both the relay link (S-R-D) and the direct link (S-D) are considered. In particular, we model the S-D, S-R, and R-D channels to be frequency-selective, and the loop-back channel at R to be of a constant attenuation and a certain delay. Under this setup, we aim to characterize the maximum achievable rate of the FD-FF relay channel by jointly optimizing the transmit power allocation over frequency at S and the frequency response of the filter at R, subject to their individual power constraints. Although this problem is non-convex, we obtain its optimal solution via the Lagrange duality method. We show that the maximum achievable rate of the considered FD-FF relay system is in fact regardless of the loop-back channel at the FD relay. This result implies that the loop-back signal, conventionally known as self-interference, is indeed not harmful to the FF-based FD relay, at least from an information-theoretic viewpoint. Furthermore, numerical results show that the proposed joint source and relay optimization achieves rate gains over other heuristic designs, and is also advantageous over the conventional approach by cancelling the relay loop-back signal as self-interference, especially when the loop-back channel gain is large (as there is a maximum interference cancellation capability at the relay).

\section{System Model}

As shown in Fig. \ref{fig:1}, there is an FD relay node R aiming to help the information transmission from a source node S to a destination node D. We consider frequency-selective channel models, where $h_{\rm SD}(t)$, $h_{\rm SR}(t)$, and $h_{\rm RD}(t)$ denote the channel impulse responses from S to D, from S to R, and from R to D, respectively, and the respective frequency responses are denoted as $H_{\rm SD}(f)$, $H_{\rm SR}(f)$, and $H_{\rm RD}(f)$, which are assumed to be time-invariant and perfectly known at all nodes for the purpose of characterizing the maximum achievable rate of the considered FD-FF relay channel. In this paper, we consider the general passband communication system over a given bandwidth with $f\in[f_c-W/2, f_c+W/2]\triangleq\Omega$, where $f_c$ and $W$ denote the central frequency and the system bandwidth, respectively.

%
%

Let $s(t)$ denote the transmitted signal by S, which is assumed to be Gaussian distributed with power spectral density $P(f)$. Also, let $r(t)$ and $x(t)$ denote the received and transmitted signals by R, respectively. Due to the short distance between the transmit and the receive antennas at R, we assume that its loop-back channel is line-of-sight with the delay $\tau > 0$ and denote the loop-back channel coefficient as $\alpha < 1$. As a result, the received signal $r(t)$ by R can be expressed as
\begin{align}
r(t) &= h_{\rm SR}(t) \otimes s(t) + \alpha x(t -\tau) + n_{\rm R}(t), \label{eqn:r}
\end{align}
where $n_{\rm R}(t)$ denotes the Gaussian noise at the receiver of R with constant power spectral density $N_0$, and $\otimes$ denotes the convolution operation.

The FD relay R operates in an analog FF manner, with the impulse response and the frequency response of the filter denoted as $\theta(t)$ and $\Theta(f)$, respectively. Here, the FF processing is considered to be implemented in the analog domain. This is for the purpose of avoiding the significant quantization error induced by the ADC if a digital filter of the same frequency response is implemented instead. Then the input-output relation between $r(t)$ and $x(t)$ at R is expressed as
\begin{align}
x(t) &= \theta(t) \otimes r(t).\label{eqn:t}
\end{align}
Note that to characterize the fundamental limits of such an analog FF design, we consider $\theta(t)$ and $\Theta(f)$ can be digitally controlled and arbitrarily designed, and ignore the physical constraints of the analog filter such as its hardware precision and processing delay.{\footnote{To practically implement such an analog FF design in frequency-selective channels with OFDM, the FF processing delay should be carefully controlled such that, at the receiver of D, the delay spread of the forwarded signal does not exceed the length of the cyclic prefix (CP). By doing so, the inter-symbol interference between consecutive OFDM symbols can be avoided.}}

By considering the signals transmitted from both S and R, the received signal at D is expressed as
\begin{align}
y(t) &= h_{\rm SD}(t) \otimes s(t) +  h_{\rm RD}(t) \otimes x(t) + n_{\rm D}(t),\label{eqn:y}
\end{align}
where $n_{\rm D}(t)$ denotes the Gaussian noise at the receiver of D with constant power spectral density $N_0$.

Based on signal input-output relations in (\ref{eqn:r}), (\ref{eqn:t}), and (\ref{eqn:y}), we have the following lemma.
\begin{lemma}\label{lemma:1}
The effective frequency response of the FD-FF relay channel between the input $s(t)$ and the output $y(t)$ is given by
\begin{align}\label{eqn:frequency:response}
H_{\rm SD}(f) + \frac{H_{\rm RD}(f)H_{\rm SR}(f)\Theta(f)}{1-\hat\alpha(f)\Theta(f)},
\end{align}
and the effective noise power spectral density at the receiver of D is
\begin{align}\label{eqn:PSD:Noise:D}
\left(\left|\frac{H_{\rm RD}(f)\Theta(f)}{1-\hat\alpha(f)\Theta(f)}\right|^2 + 1\right)N_0.
\end{align}
where $\hat\alpha(f) =\alpha \exp(-j2\pi\tau f)$. The power spectral density of the transmitted signal $x(t)$ by the FD relay R is given as
\begin{align}\label{eqn:Q:f}
Q(f) = \left|\frac{\Theta(f)}{1-\hat\alpha(f)\Theta(f)}\right|^2\left(\left|H_{\rm SR}(f)\right|^2P(f)+ N_0\right).
\end{align}
\end{lemma}
\begin{IEEEproof}
See Appendix \ref{app:A}.
\end{IEEEproof}
%

Furthermore, suppose that the power spectral densities $P(f)$ at S and $Q(f)$ at R can both vary over frequency, subject to their individual power constraints $\bar P$ and $\bar Q$, respectively:
\begin{align}
&\int_{f\in\Omega}P(f)\mathrm{d}f \le \bar{P},\label{eqn:power:S}\\
&\int_{f\in\Omega}Q(f)\mathrm{d}f  \le \bar{Q}.\label{eqn:power:R}
\end{align}

Then, we have the following theorem.
\begin{theorem}\label{theorem}
The maximum achievable rate of the frequency-selective FD-FF relay channel (in bps/Hz) is given by
\begin{align}
C = \max_{P(f)\ge 0,\Theta(f)}~~~ &\frac{1}{W}\int_{f\in\Omega}R(f)\mathrm{d}f\nonumber\\
\mathrm{s.t.}~~~&(\ref{eqn:power:S})~{\rm and}~(\ref{eqn:power:R}),\label{eqn:capacity}
\end{align}
where
\begin{align}
R(f) =  \frac{1}{2}\log_2\left(1+\frac{\left|H_{\rm SD}(f) + \frac{H_{\rm RD}(f)H_{\rm SR}(f)\Theta(f)}{1-\hat\alpha(f)\Theta(f)}\right|^2P(f)}{\left(\left|\frac{H_{\rm RD}(f)\Theta(f)}{1-\hat\alpha(f)\Theta(f)}\right|^2 + 1\right)N_0}\right).\label{eqn:Cf}
\end{align}
\end{theorem}

\begin{IEEEproof}
Based on the effective FD-FF relay channel model given in Lemma \ref{lemma:1}, together with the capacity of parallel Gaussian channels \cite{CoverBook}, this theorem can be easily proved. The details are omitted here for brevity.
\end{IEEEproof}

Note that $R(f)$ can be considered as the incremental achievable rate associated with a given frequency $f$ over the bandwidth $\mathrm{d}f$ under the given power allocation $P(f)$ at S and given frequency response $\Theta(f)$ of the filter at R. Consequently, the maximum achievable rate of the frequency-selective FD-FF relay channel can be viewed as the maximum average rate $\frac{1}{W}\int_{f\in\Omega} R(f)\mathrm{d}f$ by jointly optimizing $P(f)$ and $\Theta(f)$ over frequency, subject to the individual power constraints in (\ref{eqn:power:S}) and (\ref{eqn:power:R}) for S and R, respectively.


%

Now, it remains to solve problem (\ref{eqn:capacity}) to characterize the maximum achievable rate. However, problem (\ref{eqn:capacity}) is in general non-convex, since both its objective function and the constraint in (\ref{eqn:power:R}) are not convex. As a result, it is generally a difficult problem to solve.

\section{Joint Source and Relay Optimization}

In this section, we present the optimal solution to problem (\ref{eqn:capacity}) by first reformulating the problem and showing that its optimal solution (and thus the maximum achievable rate of the considered FD-FF relay channel) is in fact regardless of the loop-back channel coefficient $\alpha$ and delay $\tau$ at the FD relay, and then obtaining the optimal power allocation $P(f)$ at S and the filter design $\Theta(f)$ at R by using the Lagrange duality method.

\subsection{Problem Reformulation}\label{sec:III_a}

First, we define auxiliary variables $\Xi(f) = \frac{\Theta(f)}{1-\hat\alpha(f)\Theta(f)},\forall f\in\Omega$, and accordingly have
\begin{align}
\Theta(f) = \frac{\Xi(f)}{1+\hat\alpha(f)\Xi(f)},\forall f\in\Omega.\label{eqn:Theta}
\end{align}
Thanks to the one-to-one mapping between $\Xi(f)$ and $\Theta(f)$, by replacing $\Theta(f)$ as $\frac{\Xi(f)}{1+\hat\alpha(f)\Xi(f)}$, we reformulate problem (\ref{eqn:capacity}) as
{\begin{small}
\begin{align}
&\max_{P(f)\ge 0,\Xi(f)} \nonumber \\&\frac{1}{2W}\int_{f\in\Omega}\log_2\bigg(1+ \frac{\big|H_{\rm SD}(f) + H_{\rm RD}(f)H_{\rm SR}(f)\Xi(f)\big|^2P(f)}{\left(\left|H_{\rm RD}(f)\Xi(f)\right|^2 + 1\right)N_0}\bigg)\mathrm{d}f\label{eqn:obj1}\\
&\mathrm{s.t.}~\int_{f\in\Omega}\left|\Xi(f)\right|^2\left(\left|H_{\rm SR}(f)\right|^2P(f)+ N_0\right)\mathrm{d}f \le \bar{Q}\label{eqn:con2}\\
&~~~~~(\ref{eqn:power:S}).\nonumber
\end{align}
\end{small}}In problem (\ref{eqn:obj1}), it is observed that the phase of $\Xi(f)$ only affects the numerator of the signal-to-noise ratio (SNR) in the objective function, i.e., $\big|H_{\rm SD}(f) + H_{\rm RD}(f)H_{\rm SR}(f)\Xi(f)\big|^2$. As a result, to maximize the objective value, the phase of $\Xi(f)$ should align with that of $H_{\rm SD}(f)H_{\rm RD}^\dagger(f)H_{\rm SR}^\dagger(f)$, and accordingly we have
\begin{align}
\Xi(f) = \bar{\Xi}(f)\frac{H_{\rm SD}(f)H_{\rm RD}^\dagger(f)H_{\rm SR}^\dagger(f)}{\left|H_{\rm SD}(f)H_{\rm RD}^\dagger(f)H_{\rm SR}^\dagger(f)\right|},\label{eqn:Xi}
\end{align}
where $\bar\Xi(f) \ge 0$ denotes the amplitude of $\Xi(f)$, and the superscript $\dagger$ denotes the conjugate operation. Based on (\ref{eqn:Xi}), problem (\ref{eqn:obj1}) is reformulated as\vspace{-1em}

{\begin{footnotesize}\begin{align}
&\max_{P(f)\ge 0,\bar\Xi(f)\ge 0}~~~ \nonumber\\
&\frac{1}{2W}\int_{f\in\Omega}\log_2\bigg(1+ \frac{\left(\left|H_{\rm SD}(f)\right| + \left|H_{\rm RD}(f)H_{\rm SR}(f)\right|\bar\Xi(f)\right)^2P(f)}{\left(\left|H_{\rm RD}(f)\right|^2\bar\Xi(f)^2 + 1\right)N_0}\bigg)\mathrm{d}f\label{eqn:obj2}\\
&\mathrm{s.t.}~\int_{f\in\Omega}\bar\Xi(f)^2\left(\left|H_{\rm SR}(f)\right|^2P(f)+ N_0\right)\mathrm{d}f \le \bar{Q}\label{eqn:con2:2}\\
&~~~~~(\ref{eqn:power:S}),\nonumber
\end{align}
\end{footnotesize}}

\begin{remark}
It is worth noting that problem (\ref{eqn:obj2}), the reformulation of problem (\ref{eqn:capacity}) for deriving the maximum achievable rate, is irrespective of the loop-back channel coefficient $\alpha$ and delay $\tau$ at the FD relay. As a result, it can be concluded that the maximum achievable rate of the FD-FF relay channel is in fact regardless of $\alpha$ and $\tau$. This implies that the loop-back signal, conventionally known as self-interference, is indeed not harmful for the FF-based FD relay, at least from an information-theoretic viewpoint. Intuitively, this result can be explained by noting the fact that the loop-back signal is indeed a  delayed version of the output signal of the FD relay R, which contains useful information signal from S and does not change the SNR at both R and D at each frequency.
\end{remark}

\subsection{Optimal Solution to the Reformulated Problem (\ref{eqn:obj2})}\label{sec:III_b}

Now, to find the optimal solution to problem (\ref{eqn:capacity}), it is equivalent to solving problem (\ref{eqn:obj2}). Although problem (\ref{eqn:obj2}) is still non-convex in general, one can verify that it satisfies the ``time-sharing condition'' defined in \cite{YuLui2006}, for which the proof is similar to \cite{YuLui2006} and thus is omitted for brevity. The time-sharing condition implies that strong duality holds between problem (\ref{eqn:obj2}) and its Lagrange dual problem, and therefore, we can apply the Lagrange duality method to solve problem (\ref{eqn:obj2}) optimally.

Let $\mu \ge 0$ and $\lambda \ge 0$ denote the dual variables associated with the individual power constraints in (\ref{eqn:power:S}) and (\ref{eqn:con2:2}), respectively. Then the (partial) Lagrangian of problem (\ref{eqn:obj2}) is expressed as\vspace{-1em}

{\begin{footnotesize}\begin{align}
&\mathcal L(P(f),\bar \Xi(f),\mu,\lambda) = \nonumber\\
& \frac{1}{2W}\int_{f\in\Omega}\log_2\bigg(1+ \frac{\left(\left|H_{\rm SD}(f)\right| + \left|H_{\rm RD}(f)H_{\rm SR}(f)\right|\bar\Xi(f)\right)^2P(f)}{\left(\left|H_{\rm RD}(f)\right|^2\bar\Xi(f)^2 + 1\right)N_0}\bigg)\mathrm{d}f \nonumber\\&- \mu \left(\int_{f\in\Omega}P(f)\mathrm{d}f - \bar{P}\right) \nonumber\\&- \lambda\left(\int_{f\in\Omega}\bar\Xi(f)^2\left(\left|H_{\rm SR}(f)\right|^2P(f)+ N_0\right)\mathrm{d}f - \bar{Q}\right).
\end{align}\end{footnotesize}}Then the dual function of problem (\ref{eqn:obj2}) is expressed as
\begin{align}\label{eqn:obj2:dual:function}
\Upsilon(\mu,\lambda) = \max_{P(f)\ge 0,\bar \Xi(f)\ge 0} & \mathcal L(P(f),\bar \Xi(f),\mu,\lambda).
\end{align}
Accordingly, the dual problem is given as
\begin{align}
\min_{\mu\ge0,\lambda\ge0}~\Upsilon(\mu,\lambda).\label{eqn:obj2:dual}
\end{align}
Thanks to the strong duality between problem (\ref{eqn:obj2}) and its dual problem (\ref{eqn:obj2:dual}), in the following we solve problem (\ref{eqn:obj2}) by first obtaining $\Upsilon(\mu,\lambda)$ under any given $\mu\ge 0$ and $\lambda \ge 0$  via solving problem (\ref{eqn:obj2:dual:function}), and then searching over $\mu \ge 0$ and $\lambda \ge 0$ to minimize $\Upsilon(\mu,\lambda)$.

\subsubsection{Obtaining $\Upsilon(\mu,\lambda)$ via Solving Problem (\ref{eqn:obj2:dual:function})}

First, consider problem (\ref{eqn:obj2:dual:function}) under given $\mu\ge 0$ and $\lambda \ge 0$. By dropping the constant terms, problem (\ref{eqn:obj2:dual:function}) can be decomposed into a series of subproblems as follows, each corresponding to a frequency $f\in \Omega$. Note that for notational convenience, we have omitted the frequency index $f$ in the rest of this subsection unless otherwise stated.
\begin{align}
\max_{P\ge 0,\bar\Xi\ge 0} & \frac{1}{2W}\log_2\bigg(1+ \frac{\left(\left|H_{\rm SD}\right| + \left|H_{\rm RD}H_{\rm SR}\right|\bar\Xi\right)^2P}{\left(\left|H_{\rm RD}\right|^2\bar\Xi^2 + 1\right)N_0}\bigg) \nonumber\\&- \mu P  - \lambda\bar\Xi^2\left|H_{\rm SR}\right|^2P\label{eqn:f}
\end{align}

Although problem (\ref{eqn:f}) is non-convex, we find its optimal solution, denoted as $P^\star$ and $\bar\Xi^\star$, by first obtaining the optimal $P$ under given $\bar\Xi\ge 0$ and then finding the optimal $\bar\Xi$. Define
\vspace{-0.5em}

{\begin{footnotesize}\begin{align}
\beta(\bar\Xi) &\triangleq \frac{\left(\left|H_{\rm SD}\right| + \left|H_{\rm RD}H_{\rm SR}\right|\bar\Xi\right)^2}{\left(\mu  + \lambda\bar\Xi^2\left|H_{\rm SR}\right|^2\right){\left(\left|H_{\rm RD}\right|^2\bar\Xi^2 + 1\right)N_0}} \label{eqn:beta}\\
\chi(\bar\Xi)&\triangleq\bigg(\frac{1}{(2\ln2) W(\mu  + \lambda\bar\Xi^2\left|H_{\rm SR}\right|^2)} - \frac{\left(\left|H_{\rm RD}\right|^2\bar\Xi^2 + 1\right)N_0}{\left(\left|H_{\rm SD}\right| + \left|H_{\rm RD}H_{\rm SR}\right|\bar\Xi\right)^2}\bigg)^+,\label{eqn:P}
\end{align}\end{footnotesize}}where $x^+ \triangleq \max(x,0)$. Then we have the following proposition.
\begin{proposition}\label{proposition:1}
The optimal solution to problem (\ref{eqn:f}) is given as
\begin{align}\label{eqn:lemma}
\bar\Xi^{\star} & = \left\{
\begin{array}{ll}
\arg\max_{\bar\Xi\ge 0} \beta(\bar\Xi), & {\rm if}~\max_{\bar\Xi\ge 0} \beta(\bar\Xi) > (2\ln 2) W\\
0,& {\rm otherwise},
\end{array}
\right. \\
P^\star &= \chi(\bar\Xi^\star).
\end{align}
\end{proposition}
\begin{IEEEproof}
See Appendix \ref{app:B}.
\end{IEEEproof}

Note that the function $\beta(\bar\Xi)$ is quasi-concave in $\bar\Xi \ge 0$, and thus the problem $\max_{\bar\Xi\ge 0} ~\beta(\bar\Xi)$ is a quasi-convex optimization problem, for which we can obtain its optimal solution via standard convex optimization techniques such as the bisection method. Therefore, the optimal $\bar\Xi^{\star} $ in (\ref{eqn:lemma}) can be efficiently obtained. By applying Proposition \ref{proposition:1} for all frequency, problem (\ref{eqn:obj2:dual:function}) is solved with the optimal solution denoted by $P^{\star}(f)$  and $\bar\Xi^{\star}(f)$, and accordingly, $\Upsilon(\mu,\lambda)$ is obtained.
\subsubsection{Searching Optimal $\mu \ge 0$ and $\lambda \ge 0$ to Minimize $\Upsilon(\mu,\lambda)$}
Next, we search over $\mu \ge 0$ and $\lambda \ge 0$ to minimize $\Upsilon(\mu,\lambda)$. Since the dual function $\Upsilon(\mu,\lambda)$ is convex over $\mu \ge 0$ and $\lambda \ge 0$ but in general non-differentiable, we can use subgradient based methods such as the ellipsoid method \cite{BoydConvexII} to minimize it by using the fact that the subgradients of $\Upsilon(\mu,\lambda)$ are
\begin{align}
\mv \vartheta&(\mu,\lambda) = \bigg[\bar{P} - \int_{f\in\Omega}P^{\star}(f)\mathrm{d}f,\nonumber\\
&~ \bar{Q} - \int_{f\in\Omega}\bar\Xi^{\star}(f)^2\left(\left|H_{\rm SR}(f)\right|^2P^{\star}(f)+ N_0\right)\mathrm{d}f\bigg]\nonumber
\end{align}
under given $\mu$ and $\lambda$. Let the obtained optimal $\mu$ and $\lambda$ to the dual problem (\ref{eqn:obj2:dual}) be denoted as $\mu^{\star}$ and $\lambda^{\star}$. Then the optimal solution of $P^{\star}(f)$ and $\bar\Xi^{\star}(f)$ to problem (\ref{eqn:obj2:dual:function}) under $\mu^{\star}$ and $\lambda^{\star}$ is the optimal solution of problem (\ref{eqn:obj2}).

In summary, the detailed algorithm for solving problem (\ref{eqn:obj2}) is presented as Algorithm 1 in Table \ref{table:1}.

Based on the optimal $P^{\star}(f)$ and $\bar\Xi^{\star}(f)$ obtained in Algorithm 1, together with the transformations in (\ref{eqn:Theta}) and (\ref{eqn:Xi}), the optimal power allocation at S and frequency response of the filter at R are obtained as $P^{\star}(f)$  and $\Theta^{\star}(f)$, respectively.

\begin{table}[!t]\scriptsize
\caption{Algorithm for Solving Problem (\ref{eqn:obj2})}\vspace{-1em}
\label{table:framework} \centering
\begin{tabular}{|p{8cm}|}
\hline
\textbf{Algorithm 1}\\
\hline\vspace{0.01cm}
1) {\bf Initialization:} Set the iteration index $n=0$, and given an ellipsoid $\xi^{(0)} \subseteq \mathbb{R}^2$ centered at $[\mu^{(0)},\lambda^{(0)}]^T$. \\
2) {\bf Repeat:}
  \begin{itemize} \setlength{\itemsep}{0pt}
    \item[a)] Solve problem (\ref{eqn:obj2:dual:function}) under given $\mu^{(n)}$ and $\lambda^{(n)}$ by using Proposition \ref{proposition:1} to obtain $\Upsilon(\mu^{(n)},\lambda^{(n)})$;
    \item[b)] Update the ellipsoid $\xi^{(n+1)}$ based on $\xi^{(n)}$ and the subgradient $\mv \vartheta(\mu^{(n)},\lambda^{(n)})$. Set $[\mu^{(n+1)},\lambda^{(n+1)}]^T$ as the center for $\xi^{(n+1)}$.
    \item[c)] $n \gets n+1$;
  \end{itemize}
  3) {\bf Until} the stopping criteria for the ellipsoid method is met.\\
  4) {\bf Set} $\mu^{\star} = \mu^{(n)}$ and $\lambda^{\star} = \lambda^{(n)}$. Then the optimal solution $P^{\star}(f)$ and $\bar\Xi^{\star}(f)$ to problem (\ref{eqn:obj2:dual:function}) under $\mu^{\star}$ and $\lambda^{\star}$ is the optimal solution of problem (\ref{eqn:obj2}).\\
 \hline
\end{tabular}\label{table:1}\vspace{-1.5em}
\end{table}

\section{Numerical Results}

In this section, we provide numerical results to show the achievable rate of our considered FD-FF relay channel by harnessing the loop-back signal. In the parameter setting, we equally divide the bandwidth $W=10.24$ MHz into a total of $1024$ frequency sub-channels to achieve good approximation accuracy of the achievable rate. 
We consider multi-tap frequency-selective channels, 
and generate the tap coefficients of $h_{\rm SD}(t)$, $h_{\rm SR}(t)$, and $h_{\rm RD}(t)$ as Gaussian random variables with zero mean and variances $-110$ dB, $-100$ dB and $-100$ dB, respectively. This corresponds to the practical scenario when the distance between S and D is around 500 meters while R is located in the middle between them. We set the noise power at R's and D's receivers as $N_0 = -145$ dBm/Hz (which
is dominated by the receiver processing noise rather than the background thermal noise). 


\begin{figure}
\centering
 \epsfxsize=1\linewidth
    \includegraphics[width=8cm]{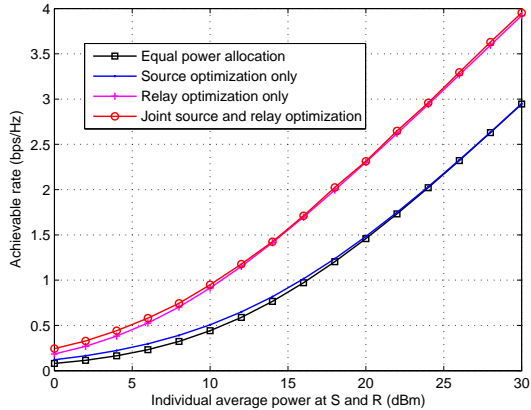}\vspace{-0em}
\caption{The achievable rate versus the individual average power at S and R assuming $\bar{P} = \bar{Q}$.
} \label{fig:2}\vspace{-1em}
\end{figure}

First, we show the achievable rate by joint source and relay optimization as compared to three heuristically designed schemes in the following. All the three heuristic schemes design the filter $\Theta(f)$ at R based on (\ref{eqn:Theta}) and (\ref{eqn:Xi}) to exploit the loop-back signal, with $\bar\Xi(f)$ specified as follows.
\begin{itemize}
  \item {\it Equal power allocation}: S and R equally allocate their individual transmit power over frequency, i.e.,
  \begin{align}
  P(f) &= \bar{P}/W,\label{eqn:P:equal}\\
  \bar\Xi(f) &= \sqrt{\bar{Q}/\big(W\big(\big|H_{\rm SR}(f)\big|^2P(f)+ N_0\big)\big)}.\label{eqn:barXi:equal}
  \end{align}
  \item {\it Source optimization only}: R equally allocates its transmit power over frequency by designing $\bar\Xi(f)$ as in (\ref{eqn:barXi:equal}), and S optimizes its transmit power $P(f)$ by solving problem (\ref{eqn:capacity}) under such a given $\bar\Xi(f)$.
  \item {\it Relay optimization only}: S equally allocates its transmit power over frequency by setting $P(f)$ as in (\ref{eqn:P:equal}), and R optimizes its $\bar\Xi(f)$ by solving problem (\ref{eqn:capacity}) under such a given $P(f)$.
\end{itemize}
Fig. \ref{fig:2} shows the achievable rate versus the individual average power at S and R by assuming $\bar{P} = \bar{Q}$. It is observed that the achievable rate of the FD-FF relay channel by joint source and relay optimization is superior to the achievable rates under the three heuristic schemes over all SNR regimes. In addition, the scheme with relay optimization only is observed to outperform that with source optimization only and perform close to the joint source and relay optimization, especially in the high-SNR regime. This is due to the fact that under our considered scenario, the relay link is much stronger than the direct link and thus the relay optimization is of more significance on the system performance.

\begin{figure}
\centering
 \epsfxsize=1\linewidth
    \includegraphics[width=8cm]{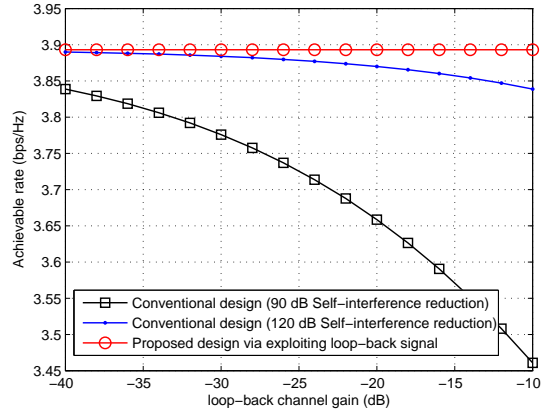}\vspace{-0em}
\caption{The achievable rate versus the loop-back channel gain $\alpha^2$ at R with $\bar{P} = \bar{Q} = 30$ dBm.} \label{fig:3}\vspace{-1em}
\end{figure}

Next, we compare the achievable rate of our considered FD-FF relay channel via exploiting the loop-back signal with that of the conventional design via self-interference cancellation at the FD relay. In the conventional design, certain residual self-interference exists after the cancellation due to e.g. the limited dynamic range of the ADC. To model this in practice, we consider two types of FD relay with self-interference reductions of $\zeta = 90$ dB and $\zeta = 120$ dB, respectively. The residual self-interference after cancellation is assumed to be white Gaussian with power spectral density given by $(\bar{Q}\alpha^2)/(W\zeta)$, which is independently added to the relay noise $n_{\rm R}(t)$ in (\ref{eqn:r}). Furthermore, we consider that in the conventional design, S and R jointly design the power allocation $P(f)$ at S and the filter $\Theta(f)$ at R by solving problem (\ref{eqn:capacity}) with $\alpha=0$.{\footnote{For the conventional design, the achievable rate maximization problem under practical non-zero residual self-interference (i.e., $\alpha > 0$) is a non-trivial problem that has not been solved in the literature yet. Therefore, for simplicity, we assume that S and R design their power allocations by assuming the self-interference can be perfectly cancelled with $\alpha = 0$. As a result, the obtained achievable rate in the conventional design is just a lower bound for comparison only.}} Fig. \ref{fig:3} shows the achievable rate versus the loop-back channel gain $\alpha^2$ at the FD relay, where the transmit powers of S and R are set as $\bar{P} = \bar{Q} = 30$ dBm. It is observed that as the value of $\alpha^2$ increases, the achievable rate of our considered FD-FF relay channel remains constant, thanks to the exploitation of the loop-back signal, while that of the conventional design is observed to decrease significantly due to the increasing  residual self-interference. Moreover, the achievable rate of our considered FD-FF relay channel substantially outperforms that under the conventional design, especially when the relay loop-back channel gain (or the residual self-interference) is large.

\section{Conclusion}

This paper studied the achievable rate of frequency-selective FD-FF relay channels. We observed that the loop-back signal is indeed a delayed version of the output signal of the relay processing which contains useful information signal of the source. By exploiting the loop-back signal, we proposed an analog FD-FF design and showed that the maximum achievable rate of such a system is in fact regardless of the loop-back channel at the FD relay, at least from an information-theoretic viewpoint. We also characterized the maximum achievable rate of this channel by jointly optimizing the transmit power allocation over frequency at the source and the frequency response of the filter at the relay, subject to their individual power constraints. Numerical results showed that the proposed joint source and relay optimization achieves rate gains over other heuristic designs, and is also advantageous over the conventional approach by cancelling the relay loop-back signal as self-interference. For the future work, it is interesting to investigate the analog FD-FF relaying design under practical OFDM systems by taking into account various issues such as relaying delay, hardware precision, etc.\vspace{-0.4em}

\appendix
\vspace{-0.4em}

\subsection{Proof of Lemma \ref{lemma:1}}\label{app:A}

First, we obtain the effective frequency response of the FD-FF relay channel between $s(t)$ and $y(t)$ by ignoring the noise $n_{\rm R}(t)$ and $n_{\rm D}(t)$ at the R's and D's receivers. In this case, we have
\begin{align}
y(t) &= h_{\rm SD}(t) \otimes s(t) +  h_{\rm RD}(t) \otimes x(t),\label{eqn:y:2}\\
r(t) &= h_{\rm SR}(t) \otimes s(t) + \alpha x(t -\tau), \label{eqn:r:2}\\
x(t) &= \theta(t) \otimes r(t).\label{eqn:t:2}
\end{align}
From (\ref{eqn:r:2}) and (\ref{eqn:t:2}), it follows that
\begin{align}
x(t)\otimes (\delta(t) - \alpha \theta(t-\tau)) = \theta(t)\otimes h_{\rm SR}(t) \otimes s(t),\label{eqn:x:s:2}
\end{align}
where $\delta(t)$ denotes the Dirac delta function.

Note that the cross-correlation of real functions $y(t)$ and $s(t)$ is equivalent to the convolution of $y(-t)$ and $s(t)$. Therefore, by denoting $R_{ys}(t)$ as the cross-correlation of $y(t)$ and $s(t)$, we have $R_{ys}(t) = y(-t) \otimes s(t)$. By using this fact together with (\ref{eqn:y:2}) and (\ref{eqn:x:s:2}), it follows that
\begin{align}
&R_{ys}(t)\otimes (\delta(-t) - \alpha \theta(-t-\tau)) \nonumber\\
=&R_{ss}(t)\otimes (h_{\rm SD}(-t) \otimes (\delta(-t) - \alpha \theta(-t-\tau) \nonumber\\
 &+ h_{\rm RD}(-t) \otimes \theta(t)\otimes h_{\rm SR}(t)),\label{eqn:time}
\end{align}
where $R_{ss}(t)$ denotes the autocorrelation of $s(t)$. By taking the Fourier transformation of (\ref{eqn:time}), we can have
\begin{align}
\frac{S_{ys}(f)}{S_{ss}(f)} = H_{\rm SD}(f) + \frac{H_{\rm RD}(f)H_{\rm SR}(f)\Theta(f)}{1-\hat\alpha(f)\Theta(f)},
\end{align}
where $S_{ys}(f)$ and $S_{ss}(f)$ denote the Fourier transforms of $R_{ys}(t)$ and $R_{ss}(t)$, respectively. As a result, it follows from \cite[Chapter 10.2]{PowerSpectralDesnsity} that the effective frequency response of the FD-FF relay channel between the input $s(t)$ and the output $y(t)$ is obtained as in (\ref{eqn:frequency:response}).

Next, we obtain the power spectral density of the effective noise at the D's receiver by deriving $S_{yy}(f)$ (i.e., the Fourier transform of $R_{yy}(t)$) via setting the input signal $s(t)$ to be zero. In this case, we have
\begin{align}
x(t)\otimes (\delta(t) - \alpha \theta(t-\tau)) &= \theta(t) \otimes n_{\rm R}(t),\\
y(t) &=  h_{\rm RD}(t) \otimes x(t) + n_{\rm D}(t).
\end{align}
After some manipulations, we have
\begin{align}
&R_{yy}(t)\otimes (\delta(-t) - \alpha \theta(-t-\tau))\otimes (\delta(t) - \alpha \theta(t-\tau)) \nonumber\\
=&R_{n_{\rm R}n_{\rm R}}(t) \otimes h_{\rm RD}(-t) \otimes \theta(-t)\otimes h_{\rm RD}(t) \otimes \theta(t) \nonumber\\
& + R_{n_{\rm D}n_{\rm D}}(t) \otimes(\delta(-t) - \alpha \theta(-t-\tau))\otimes (\delta(t) - \alpha \theta(t-\tau)).\label{eqn:time:NoisePSD}
\end{align}
By taking the Fourier transformation of (\ref{eqn:time}) and noting that $S_{n_{\rm R}n_{\rm R}}(f) = S_{n_{\rm D}n_{\rm D}}(f) = N_0$, we obtain $S_{yy}(f)$  as given in (\ref{eqn:PSD:Noise:D}), which is the power spectral density the effective noise at the D's receiver.

Finally, we obtain the power spectral density of the transmitted signal $x(t)$ by R. It follows from (\ref{eqn:r}) and (\ref{eqn:t}) that
\begin{align}
x(t)\otimes (\delta(t) - \alpha \theta(t-\tau)) &= \theta(t) \otimes (h_{\rm SR}(t) \otimes s(t) + n_{\rm R}(t)).\label{eqn:x:s}
\end{align}
Based on (\ref{eqn:x:s}), we have
\begin{align}
&R_{xx}(t) \otimes  (\delta(-t) - \alpha \theta(-t-\tau))\otimes (\delta(t) - \alpha \theta(t-\tau)) \nonumber\\
=& R_{ss}(t) \otimes (\theta(-t)\otimes h_{\rm SR}(-t)) \otimes (\theta(t)\otimes h_{\rm SR}(t)) \nonumber\\
&+R_{n_{\rm R}n_{\rm R}}(t)\otimes\theta(-t)\otimes \theta(t).\label{eqn:R_xx}
\end{align}
By taking the Fourier transformation of (\ref{eqn:R_xx}) and using the fact that $S_{ss}(f) = P(f)$ and $R_{n_{\rm R}n_{\rm R}}(t) = N_0$, we can have the power spectral density $Q(f) = S_{xx}(f)$ as given in (\ref{eqn:Q:f}). Therefore, this lemma is finally proved.

\subsection{Proof of Proposition \ref{proposition:1}}\label{app:B}

First, suppose that $\bar\Xi$ is given. In this case, the objective value of problem (\ref{eqn:f}) is concave over $P\ge 0$, and as a result, we can obtain its optimal maximizer as $P = \chi(\bar\Xi)$ with $\chi(\cdot)$ given in (\ref{eqn:P}). Accordingly, the resulting objective value of (\ref{eqn:f}) is given as
\begin{align}
v(\bar\Xi) = & \bigg(\frac{1}{W}\log_2\left(\frac{\beta(\bar\Xi)}{(2\ln2) W}\right)\bigg)^+  - \bigg(\frac{1}{(2\ln2) W} - \frac{1}{\beta(\bar\Xi)}\bigg)^+.\label{eqn:vp}
\end{align}

Next, with $\chi(\bar\Xi)$ in (\ref{eqn:P}) at hand, finding the optimal $\bar\Xi^\star$ to problem (\ref{eqn:f}) is equivalent to finding the optimal $\bar\Xi \ge 0$ to maximize $v(\bar\Xi)$, i.e.,
\begin{align}\label{eqn:f:2}
\bar\Xi^\star = \arg \max_{\bar\Xi\ge 0} ~v(\bar\Xi).
\end{align}
We solve problem (\ref{eqn:f:2}) by considering the two cases with $\max_{\bar\Xi\ge 0} \beta(\bar\Xi) \le (2\ln 2) W$ and $\max_{\bar\Xi\ge 0} \beta(\bar\Xi) > (2\ln 2) W$, respectively.

First, consider $\max_{\bar\Xi\ge 0} \beta(\bar\Xi) \le (2\ln 2) W$. In this case, it follows that $\beta(\bar\Xi) \le (2\ln 2) W, \forall \bar\Xi \ge 0$. As a result, we have $v(\bar\Xi) = 0,\forall \bar\Xi \ge 0$. Therefore, the optimal solution $\bar\Xi^{\star}$ to problem (\ref{eqn:f:2}) can be any non-negative value, and we set $\bar\Xi^{\star} = 0$ in (\ref{eqn:lemma}) without loss of optimality.

Next, consider $\max_{\bar\Xi\ge 0} \beta(\bar\Xi) > (2\ln 2) W$. Note that when $\beta(\bar\Xi) > (2\ln 2) W$, the function $v(\bar\Xi)$ is re-expressed as
\begin{align}
\bar v(\bar\Xi) = & \frac{1}{2W}\log_2\left(\frac{\beta(\bar\Xi)}{(2\ln2) W}\right)  - \frac{1}{(2\ln2) W} + \frac{1}{\beta(\bar\Xi)} > 0,\label{eqn:vp2}
\end{align}
which is monotonically increasing over $\beta(\bar\Xi)$ if $\beta(\bar\Xi) > (2\ln 2) W$. In this case, $\bar\Xi$ that maximizes $\beta(\bar\Xi)$ is also the maximizer for $v(\bar\Xi)$. As a result, we have $\bar\Xi^{\star} = \arg\max_{\bar\Xi\ge 0} \beta(\bar\Xi)$ in (\ref{eqn:lemma}).

By combining the above two cases, we obtain $\bar\Xi^{\star}$ in (\ref{eqn:lemma}). Furthermore, substituting $\bar\Xi^{\star}$ into $P = \chi(\bar\Xi)$, the optimal $P^\star$ in (\ref{eqn:P}) follows. Therefore, this proposition is proved.

%


\end{document}